\newif\iffigs\figstrue
\documentclass[12pt]{article}
\usepackage{latexsym,amssymb}
  \iffigs
   \input{epsf}
\else
\fi

\newcommand{\eq}{\begin{equation}}
\newcommand{\eqa}{\begin{eqnarray}}
\newcommand{\en}{\end{equation}}
\newcommand{\ena}{\end{eqnarray}}
\newcommand{\enn}{\nonumber \end{equation}}

\def\sk{\vskip .4cm}
\def\noi{\noindent}
\def\om{\omega}

\def\al{\alpha}
\def\la{\lambda}

\def\Ga{\Gamma}

\def\epsi{\varepsilon}
\def\we{\wedge}

\def\de{\delta}

\def\part{\partial}

\def\n2{{{N+1} \over 2}}

\def\square{{\,\lower0.9pt\vbox{\hrule \hbox{\vrule height 0.2 cm
\hskip 0.2 cm \vrule height 0.2 cm}\hrule}\,}}

\def\Q.E.D.{\rightline{$\Box$}}

\def\dti{{\tilde d}}

\def\omb{{\bf \mbox{\boldmath $\om$}}}

\newcommand{\ul}{\underline}
\newcommand{\bea}{\begin{eqnarray}}
\newcommand{\eea}{\end{eqnarray}}
\newcommand{\beq}{\begin{equation}}
\newcommand{\eeq}{\end{equation}}

\begin{document}
\begin{titlepage}
\vskip -1cm \rightline{DFTT-11/2001}
\rightline{April 2001} \vskip 1em
\begin{center}
{\large\bf Supersymmetric domain wall $\times$ $G/H$ solutions of
$IIB$ supergravity }
\\[2em]
{\bf L. Castellani ${}^{1,2}$ and L. Sommovigo ${}^2$}
\\[1em] ${}^1${\sl Dipartimento di
Scienze e Tecnologie Avanzate,
 Universit\`a del Piemonte Orientale, Italy;} \\ ${}^2$ {\sl
Dipartimento di Fisica Teorica and Istituto Nazionale di Fisica
Nucleare\\ Via P. Giuria 1, 10125 Torino, Italy.} \\ {\small
castellani@to.infn.it, sommovigo@to.infn.it}\\[2em]
\end{center}
\vskip 4 cm
\begin{abstract}
\sk

 1-brane nonmaximally supersymmetric solutions of $D=10$
chiral supergravity are discussed. In the dual frame, their near
brane geometry is the product of a 3-dimensional domain wall
spacetime and a 7-dimensional homogeneous Einstein space
$(G/H)_7$.

\end{abstract}

\begin{center}
\vskip 6cm
 \hrule
  \vskip .2cm
   \noi  {\footnotesize
  Supported in part by EEC under RTN contract HPRN-CT-2000-00131}
  \end{center}

\end{titlepage}
\newpage
\setcounter{page}{1}


It has been conjectured some time ago \cite{dwqft1,dwqft2,dwqft3}
that the $AdS/CFT$ correspondence (for a review see \cite{adscft}
and references therein) could be extended to a $DW/QFT$
correspondence between supergravity on domain wall spacetimes and
(nonconformal) quantum field theories living on their boundary.
This is a generalization of the gauge/gravity correspondence, well
tested in the framework of $AdS/CFT$, but still in need of
explicit support in the nonconformal scenarios.

In the present Letter we study in detail a particular class of
1-brane solutions of $D=10$ IIB chiral supergravity, and classify
their supersymmetry content. These solutions involve nontrivial
scalar and two-form fields. In the dual frame of ref.
\cite{dwqft2} (obtained from the Einstein frame after a Weyl
transformation on the metric) the near brane geometry factorizes
into the product of a three dimensional domain wall spacetime
$(DW)_3$ and a compact homogeneous $7$-dimensional coset manifold
$(G/H)_7$. These solutions preserve a fraction $N/16$ of the
original $D=10$ supersymmetry, where $N$ is the number of Killing
spinors of $G/H$. The literature on $p$-brane solutions of
supergravity theories is quite extensive, see for ex. the review
\cite{stelle}.

Our motivation is to set up a convenient playing ground for
testing the $DW/QFT$ correspondence between $d=3$ nonmaximal
supergravities and $d=2$ nonconformal field theories.

Compact (and noncompact) gaugings of maximal supergravities in
$d=3$ have been recently constructed in \cite{nicolai}. It will be
worthwhile to investigate also the nonmaximal cases: some of these
should correspond to consistent reductions of the $D=10$ solutions
we are discussing. For a recent review on supergravity gaugings in
diverse dimensions and their use in $p$-brane physics see for ex.
\cite{fre2001}.

Chiral IIB $D=10$ supergravity contains a complex anti-Weyl
gravitino $\psi_{M}$ and a complex Weyl spinor $\lambda$. The
bosonic fields are the graviton $g_{MN}$, a complex antisymmetric
tensor $A_{MN}$, a real antisymmetric tensor $A_{MNRS}$
(restricted by a self-duality condition) and a complex scalar
$\phi$.

After setting the spinor fields to zero, the field equations read
\cite{js2b,hw,cp}:
 \eqa
 & &2 R_{MN}=P_M P_N^* + P_M^* P_N + {1 \over 6}
 F^{PQRS}_{~~~~~~M} F_{PQRSN} + \nonumber \\
 & & ~~~~~~~~~~~~~~+{1\over 8} ( G^{PQ}_{~~~M} G^*_{PQN}
 + G^{*~PQ}_{~~~~~M} G_{PQN} - {1\over 6} g_{MN} G^{PQR}
 G^*_{PQR}) \label{einsteineq}\\
 & & F_{M_1-M_5}={1 \over 5!} \epsilon_{M_1-M_5N_1-N_5}
 F^{N_1-N_5} \label{Feq}\\
 & & (\nabla^S - i Q^S)G_{MNS}=P^S G^*_{MNS}-{2i\over 3} F_{MNPQR}
 G^{PQR} \label{Geq}\\
 & & (\nabla^M - 2i Q^M) P_M= - {1 \over 24} G^{PQR} G_{PQR}
 \label{scalareq}
 \ena
where  $R_{MN} \equiv R^S_{~M~SN}$ and the curvature two-form is
defined as $R^S_{~M} \equiv dB^S_{~M}+B^S_{~N}\we B^N_{~M}$; the
vectorial quantities $P_M$ (complex) and $Q_M$ (real) are related
to the scalar field and its first derivative, $F_{PQRSN}$ and
$G_{PQN}$ to the field strengths of the four-form and of the
two-form, and $\nabla$ is the Lorenz covariant derivative. Here we
have adopted the normalizations and conventions of  \cite{js2b};
note however a sign correction in (\ref{scalareq}), already found
in \cite{cp} and noted also in \cite{pw}. Moreover the following
Bianchi identities hold (a consequence of the field definitions):
\eqa
 & & (\nabla_{[M} - 2iQ_{[M})P_{N]}=0,~~~\partial_{[M} Q_{N]}=
-i P_{[M} P^*_{N]}\label{bianchipq}\\ & &(\nabla_{[M} -i
Q_{[M})G_{NRS]}=-P_{[M} G^*_{NRS]}\label{bianchig}\\ &
&\partial_{[N} F_{M_1..M_5]}={1 \over 8} \mbox{Im} ~G_{[NM_1M_2}
G_{M_3M_4M_5]}^* \label{bianchif}
 \ena

 The supersymmetry variations of the
bosonic fields are proportional to Fermi fields, and  these vanish
in the type of backgrounds we are considering. On the other hand,
the supersymmetry variations of the fermionic fields are:
 \eqa
 & & \de \la = i \Ga^M \epsi^* P_M - {1\over 24} i G_{MNP} \Ga^{MNP}
 \epsi \label{supersymla} \\
 & & \de \psi_M = (\nabla_M - {i\over 2} Q_M) \epsi +
 {i \over 480} F_{N_1-N_5}\Ga^{N_1-N_5} \Ga_M \epsi  + \nonumber\\
 & &~~~~~+ {1 \over 96} (\Ga_M^{~N_1-N_3} G_{N_1-N_3} - 9
 \Ga^{N_1N_2} G_{MN_1N_2}) \epsi^* \label{supersympsi}
 \ena
(in backgrounds with $\psi=0, \la=0$) \cite{js2b}. A solution is
 supersymmetric if there exist spinors $\epsi$ for which
these variations vanish in the corresponding background.

We search for solutions of the IIB field equations of the type
(two-blocks brane Ansatz):

\bea & ds^2=e^{2A(r)}dx^\mu dx^\nu
\eta_{\mu\nu}-e^{2B(r)}[dr^2+r^2
\lambda^{-2}ds_{G/H}^2]\label{ans1}\\ &
A_{\mu\nu}=\varepsilon_{\mu\nu} e^{C(r)}\label{ans2}\\ &
P_{\bullet}=-E' (r)\label{ans3}\\ & Q_M=
F_{M_1...M_5}=0\label{ans4}\eea

\noindent where $\lambda$ is a constant parameter with dimension
of a length, the metric is
$\eta_{\ul{\mu}\ul{\nu}}=(+,-,\dots,-)$, and $A$, $B$, $C$, $E$
are real functions of $r$. The indices run as follows: $\mu,\nu
\dots=1,2$; $\bullet$ labels the radial direction; $m,n \dots $
run on the internal coset manifold $G\over{H}$ directions;
 $M, N\dots$ run over $1...10$.
 The 3-form $G_{\mu\nu\rho}$ is proportional to the curl of the
potential $A_{\mu\nu}$:

\beq G_{\mu\nu\rho} = 3 e^{E(r)} \partial_{[\mu} A_{\nu\rho]}\eeq

Eq.s (\ref{ans1})-(\ref{ans4})
  provide the
 $G/H$ generalization of the standard (two-blocks) $p$-brane
 Ansatz extensively considered in the literature (see for instance
 \cite{stelle}). As such, we know that it satisfies the
  field eq.s (\ref{einsteineq})-(\ref{scalareq}) if the
  functions $A,B,C,E$ have a specific form, and the internal coset
  space is an Einstein manifold \cite{duff1995,GHbrane}. This we now verify in detail:
 in fact the explicit expression for the connection will be useful
 when discussing the supersymmetry content of the solutions.

The vielbein and spin connection corresponding to the Ansatz are
given by:

\bea &
E^{\ul{\mu}}=e^{A(r)}dx^{\mu},~~~E^{\ul{\bullet}}=e^{B(r)}dr,~~~E^{\ul{m}}=e^{B(r)}
r \lambda^{-1}\mathbf{E}^{\ul{m}}\nonumber\\ &
\omega^{\ul{\mu}\ul{\nu}}=0,~~\omega^{\ul{\mu}\ul{\bullet}}=-e^{-B(r)}A'
(r)E^{\ul{\mu}},~~ \omega^{\ul{\mu}\ul{m}}=0\nonumber\\ &
\omega^{\ul{m}\ul{\bullet}}=-e^{-B(r)}(B' (r)+r^{-1})E^{\ul{m}},~~
\omega^{\ul{m}\ul{n}}= \omb^{\ul{m}\ul{n}}\nonumber\eea

\noindent where flat indices are underlined, and
$\mathbf{E}^{\ul{m}}$, $\omb^{\ul{m}\ul{n}}$ are the vielbein and
spin connection of the coset manifold $G/H$.\\

 \noindent It is convenient to express the only nonvanishing
 components of the 3-form field strength using flat indices:

\beq G_{\ul{\mu}\ul{\nu}\ul{\bullet}}=C' e^{C+E-2A-B}
\varepsilon_{\ul{\mu}\ul{\nu}} \label{ans5} \eeq

\noi The Ricci tensor is:

\bea & R_{\ul{\mu}\ul{\nu}}= {1\over{2}} \eta_{\mu \nu} e^{-2B}
(A'' + 2{A'}^2 +6A' B' +7 r^{-1}A' )\label{ricci1}\\ &
R_{\ul{\bullet}\ul{\bullet}}=-{1\over{2}}e^{-2B}(2A'' + 2{A'}^2
-2A' B' +7B'' +7r^{-1} B' )\label{ricci2}\\ &
R_{\ul{m}\ul{n}}={1\over{2}}\eta_{mn}e^{-2B}(2A' B' +2r^{-1} A'
+13r^{-1} B' +B'' +6{B'}^2 +6r^{-2}) +
\mathbf{R}_{\ul{m}\ul{n}}\label{ricci3}\eea

\noindent where $\mathbf{R}_{\ul{m}\ul{n}}$ is the Ricci tensor of
the coset manifold.

Inserting the Ansatz (\ref{ans1})-(\ref{ans4}) and (\ref{ans5})
into the Einstein field eq. (\ref{einsteineq}) yields

\bea & 2R_{\ul{\mu}\ul{\nu}}= {3\over{8}} \eta_{\mu\nu}{C'}^2
e^{2(C+E-2A-B)}\label{einstein1}\\ & 2R_{\ul{\bullet}\ul{\bullet}}
= - {3\over{8}} {C'}^2 e^{2(C+E-2A-B)} + 2
{E'}^2\label{einstein2}\\ & 2 R_{\ul{m}\ul{n}} = - {1\over{8}}
\eta_{\ul{m}\ul{n}} {C'}^2 e^{2(C+E-2A-B)}\label{einstein3}\eea

\noindent while equations (\ref{Geq}) and (\ref{scalareq}) become:

\bea  & C'' + {C'}^2 - 2 A'C' +6B'C' +7 r^{-1} C' +2
C'E'=0\label{divergence}\\ & E'' + 6 E'B' +2 A'E' +7 r^{-1}E' +
{1\over{4}} {C'}^2 e^{2(C+E-2A)}=0\label{scalarequation}\eea

The Ansatz satisfies the Bianchi identities for $Q$ and $F$
trivially.  The other two Bianchi identities hold because of the
particular form of the spin connection (for ex. use
$\om_{[\mu~~\nu]}^{~~\bullet}=0=\om_{[m~~n]}^{~~\bullet}$ in the
$P$ Bianchi identity).
 \sk

Requiring that the solutions preserve a certain amount of the
$D=10$ supersymmetry, i.e. that the variations
(\ref{supersymla}),(\ref{supersympsi}) vanish, places further
restrictions on the functions $A,B,C,E$, which we now deduce. We
adopt the following real representation for the $SO(1,9)$ gamma
matrices ($32 \times 32$):

\beq \mathbf{\Gamma}^M= \left( \gamma^{\mu}\otimes \mathbf{1_8}
\otimes \sigma^2 , \gamma^{\bullet} \otimes \mathbf{1_8} \otimes
\sigma^2,\mathbf{1_2} \otimes \Gamma^{m} \otimes \sigma^1
\right)\eeq

\noindent where the $SO(1,2)$ $\gamma$-matrices are
$\left(\gamma^1,\gamma^2,\gamma^{\bullet}\right) =
\left(\sigma^2,i\sigma^1,i\sigma^3 \right)$ and $\sigma^i$ are the
Pauli matrices. $\Gamma^{m}$ are the real $SO(7)$ matrices given
by the octonionic structure constants (see for ex. \cite{ads3GH}).
This tensor representation is well adapted to our (2+1+7) Ansatz.

Correspondingly, any anti-Weyl  $D=10$ spinor $\epsilon$ can be
decomposed as:

\beq \epsilon=c~\xi \otimes \eta(r,y) \otimes \left(
\begin{array}{c} 0\\ 1  \end{array} \right) \eeq

\noindent where $\xi$ is a constant real $SO(1,2)$ spinor, $\eta$
 a real $SO(7)$  spinor, and $c \in \mathbb{C}$.

Requiring that the supersymmetry variation (\ref{supersymla})
vanishes in the Ansatz background is equivalent to:

\eqa {1\over{4}} \alpha~ C' e^{C+E-2A} = E' \label{supersym11} \\
\left( \mathbf{1_2}- \sigma^3 \right)\xi = 0 \label{supersym12}
\ena

\noindent where $\alpha = {c\over{c^*}} = \pm 1$ (so that $\alpha
=\alpha^{-1}$); note that eq.(\ref{supersym12}) projects out half
of the components of the $d=3$ supersymmetry parameter. Using the
condition on $\xi$, the supersymmetry variation $\de \psi_{\mu}$
of the gravitino vanishes provided

\beq A' + {3\over{8}} \alpha~ C'
e^{C+E-2A}=0\label{supersym21}\eeq

\noindent Together with (\ref{supersym11}), this relation gives
$A$ and $C$ in terms of $E$, and fixes $\al$:

\eq  A=-{3\over{2}}E,~~~ C=-4E,~~~ \alpha=-1 \label{ace}\en

\noindent Using these equalities in the condition $\de
\psi_{\bullet}=0$ yields:

\beq \partial_r \eta = -{1\over{4}} A' \eta \eeq

\noindent whose solution is:

\beq \eta(r,y) = \eta_0 (y)~ e^{-{1\over{4}}A} \eeq

\noindent Finally $\de \psi_m=0$ leads to:

\bea & B' - {1\over{8}} \al ~ C' e^{C+E-2A} =0
\label{supersym22}\\ & \left( \nabla_m^{G/H} +
{1\over{2\lambda}}\Gamma_m \right) \eta_0 =0 \eea

\noindent where $\nabla_m^{G/H}$ is the $G/H$-covariant
derivative.  From the first equation we find $B$ in terms of $E$:

\beq B= {1\over{2}}E \label{be}\eeq

\noindent while the second, after identifying the Freund-Rubin
parameter as $e={1\over{2\lambda}}$, reduces to the familiar
Killing spinor equation for a $G\over{H}$ found in studying the
residual supersymmetry of Freund-Rubin vacua. The solutions of
this equation have been exhaustively studied in the eighties, see
for ex. \cite{crw,dnp,cdf} for a review.
 \sk
 Now we come back to
the equations of motion.  After expressing the functions $A,B,C$
in terms of $E$, the four equations (\ref{einstein1}),
(\ref{einstein2}), (\ref{divergence}) and (\ref{scalarequation})
all reduce to:

\beq E''+{7\over r}E' + 4 (E')^2 =\nabla^2 E + 4 {E'}^2 =0
\label{laplace}\eeq

\noindent or
 \eq
  \nabla^2 e^{4E(r)}=0
\en
\noi  where $\nabla^2$ is the $(G/H)_7$-covariant Laplacian.

This equation is solved by:
 \eq
  e^{4E(r)}= H(r) \equiv 1 + {k\over{r^6}}
 \en
\noi and we have chosen the integration constant such that
$E(\infty)=0$. With use of (\ref{ace}),(\ref{be}) we determine the
exponentials appearing in the Ansatz:

\eq e^{2A} = H(r)^{-{3\over{4}}},~~~ e^{2B} = H(r)^{1\over{4}},~~~
e^C = H(r)^{-1} \en

Finally, the Einstein equation in the $G/H$ directions
(\ref{einstein3}) becomes:

\beq \mathbf{R}_{mn}= - {3\over{\lambda^2}}\mathbf{g}_{mn} \eeq

\noindent where $\mathbf{g}_{mn}$ is the $G$-invariant $G/H$
metric. This equation is solved by any compact 7-dimensional
Einstein manifold, and in particular by the homogeneous manifolds
$(G/H)_7$ endowed with a $G$-invariant Einstein metric, classified
in \cite{crw}. A short review on coset space geometry can be found
in \cite{cdf,cGH}. For self-consistency we recall here the subset
of such manifolds that admit $N \geq 1$ Killing spinors. The
notations are as in ref.s \cite{crw,cdf}.

\begin{table}[ht]
\begin{center}
\caption{7-dimensional Einstein coset spaces with Killing spinors}
\label{table1}
 \vskip .3cm
 \begin{tabular}{|c|c|c|c|}\hline
  G/H &   $G$     &   $H$   &   $N$    \\
 \hline
 \hline
{}&{}&{}&{}\\
 $S^7$ &  $SO(8)$ &   SO(7) &   8             \\
 \hline
 squashed $S^7$ & $SO(5) \times SO(3)$ & $ SO(3)\times SO(3)$& 1\\
 \hline
 $M^{ppr}$ & $SU(3) \times SU(2) \times U(1)$ & $SU(2) \times
U(1)^2$ & 2\\ \hline $N^{010}$ & $SU(3) \times SU(2)$ & $
SU(2)\times U(1)$& 3\\ \hline $N^{pqr}$ & $SU(3) \times U(1)$ &
$U(1)^2$& 1\\ \hline $Q^{ppp}$ & $SU(2)^3 $ & $U(1)^3$& 2\\ \hline
$B_{irred}^7$ & $SO(5)$ & $ SO(3)_{max}$& 1\\ \hline $V_{5,2}$ &
$SO(5)) \times U(1)$ & $ SO(3)\times U(1)$& 2\\ \hline
 \end{tabular}
 \end{center}
\end{table}

 To each of these coset spaces $G/H$ corresponds a supersymmetric,
 nonconformal 1-brane solution of IIB supergravity, where
 $N/16$ of the $D=10$ spacetime supersymmetries are preserved.
 For example the round seven sphere solution has $1/2$ of the
 original $D=10$ supersymmetry. This halving of supersymmetry is a
 familiar phenomenon in brane solutions \cite{stelle}; technically here it arises
 because of eq. (\ref{supersym12}).

 These 1-brane solutions of IIB supergravity are a special case of the p-brane
 classical solutions of the D-dimensional action
 \eq
 \int d^D x \sqrt{-g} \left[ 2 R[g] -{1\over 2}(\part \phi)^2 -
  {1 \over 2(p+2)!} e^{-a\phi} F^2_{p+2} \right] \label{action}
  \en
corresponding to a consistent truncation of some D-dimensional
supergravity bosonic action, involving a scalar field and a $p+2$
form $F_{p+2}$. As shown in \cite{duff1995,GHbrane} the field
equations derived from (\ref{action}) admit the following
elementary (or electric) p-brane solution:
 \eqa
 ds^2 &=& H(r)^{-{4 \dti \over \Delta (D-2)}}
 dx^{\mu} \otimes dx^{\nu} \eta_{\mu\nu}
 -H(r)^{4 d \over \Delta (D-2)} (dr^2 + {r^2\over \la^2}
 ds^2_{X})\label{genmetric} \\
 F_{p+2} &=& {2 \over \sqrt{\Delta}} \epsi_{\mu_1...\mu_{p+1}}
 d[H(r)^{-1}] \we dx^{\mu_1} \we ...\we dx^{\mu_{p+1}}\\
 e^{\phi(r)} &=& H(r)^{-{2a\over \Delta}}
 \label{gensol}\ena
\noi where $x^{\mu}, (\mu=0,...p)$ are the coordinates on the
$p$-brane worldvolume, and $d \equiv p+1$, $\dti \equiv D-d-2$ are
the worldvolume dimensions of the $p$-brane and its magnetic dual.
The remaining coordinates of the $D$-dimensional spacetime span a
generalized cone with radial coordinate $r$ (radial distance from
the brane) and whose basis is a homogeneous $(D-d-1)$-dimensional
Einstein space $X$ with metric $ds_{X}^2$. Moreover
 \eqa
 & \Delta \equiv a^2 +2~ {d\dti\over D-2}\\
 & H(r) \equiv 1+{k\over r^{\dti}}
 \ena
\noi where $k$ is the electric charge of the brane, and $H(r)$ is
a harmonic function ($\nabla^2 H(r)=H''+(\dti+1)r^{-1} H'=0$) on
the space transverse to the brane worldvolume. When $X$ is the
usual round sphere, the metric (\ref{genmetric}) is asymptotic to
$D$-dimensional Minkowski spacetime as $r$ goes to infinity; if
$X$ is any other Einstein space the same metric is no longer
asymptotic to $D$-Minkowski spacetime, although it is still
asymptotically flat \cite{duff1995}.

The IIB supergravity 1-brane solution we have discussed in detail
can be easily cast in the form (\ref{gensol}) with the following
identifications
 \eq
 E(r)=-{\phi (r) \over 2},~ ~G_3=e^{-{\phi\over 2}} F_3,~ ~a=1,
  ~~d=2,~~\dti=6,~~ \Delta=4
 \en

The discussion of ref. (\cite{dwqft2}) can be applied ``in toto",
with a slight generalization: the sphere $S^{D-d-1}$ becomes an
arbitrary $G/H$ Einstein manifold, with $N \geq 1$ Killing
spinors. Then in the so called ``dual" frame the near brane
geometry factorizes into $DW_3 \times (G/H)_7$, i.e. a
3-dimensional domain-wall spacetime times the Einstein manifold
$(G/H)_7$. The reduction of the action (\ref{action}) over the
$G/H$ part of the near-brane geometry proceeds in the same fashion
as discussed in ref.s \cite{dwqft2,cllp,dwqft3}. One then obtains
a $D=3$ action admitting a domain wall spacetime with parameter
$\Delta_{DW}=-3$. What is exactly the gauged $D=3$ supergravity
whose truncation yields this $D=3$ action is still to be
elucidated. Only very recently a systematic analysis of maximal
(non)compact gaugings of $D=3$ supergravity has been carried out
in \cite{nicolai}. Clearly also the nonmaximal gaugings are of
interest for the solutions we have studied here, since these have
nonmaximal supersymmetry.

 In conclusion, we have given the detailed supersymmetry analysis
 of 1-brane solutions of IIB supergravity having in mind to find
 the corresponding noncompact gaugings of nonmaximal $D=3$
 supergravity. This remains still to be done, and to be used to
 test the $DW_3/QFT_2$ correspondence between these classical
 supergravities and the 2-dimensional QFT living on the brane
 worldsheet.

\vskip 1cm

\noi {\bf Acknowledgements}
 \sk
 \noi  It is a pleasure to acknowledge useful
discussions with Pietro Fr\'e.


\vfill\eject
\end{document}